\documentclass{article}
\usepackage{epsfig,times,textcomp,subfigure,amssymb,lineno}

\newcommand{\bal}{\begin{eqnarray*}}
\newcommand{\eal}{\end{eqnarray*}}

\begin{document}

\title{Likelihood description for comparing data with simulation of limited statistics}
\author{D.~Chirkin\\dima@icecube.wisc.edu\\Dept.~of Physics and WIPAC, University of Wisconsin, Madison, WI 53706, USA}
\maketitle

\begin{abstract}
It is often not possible to construct a probability density function that describes the data. This can happen if there is no analytic description, and the number of parameters is too large so that it is impossible to simulate and tabulate all combinations. In these situations it is still interesting to rank simulation sets performed with different parameters in how well they compare to data. We propose a solution that appears to be better suited to this task than some of the obvious alternatives.
\end{abstract}

\section{Introduction}
It is often the case that the mean rate of counts in a bin is not known exactly but rather approximated with simulation. The simulation can be repeated many times, obtaining a total number of counts of $s$ in $n_s$ trials, and the expected rate of counts is often approximated as $\mu=s/n_s$. For the sake of generality let's assume that we repeat the experiment $n_d$ times and collect a total of $d$ counts. In order to fit for some unknown property of the experiment one often maximizes the likelihood, and for convenience that is usually done by minimizing the minus log likelihood, $-\ln{\cal L}$. The minus log likelihood based on the Poisson probability of our observation is given by
\[
-\ln{\cal L} = \ln d!+n_d\mu-d\cdot\ln(n_d\mu).
\]
Unfortunately this expression is only an approximation as the quantity $\mu$ is not known precisely but was calculated from simulation and is known within statistical uncertainties corresponding to the total number of simulated counts $s$ in our bin.

When the counts $s$ in simulation and $d$ in data are large, one minimizes the $\chi^2$:
\[
\chi^2 = {(s/n_s-d/n_d)^2\over s/n_s^2+d/n_d^2},
\]
where the total uncertainty in the denominator is computed as the square root of the sum of squares of the mutually independent statistical uncertainties of $s/n_s$ and $d/n_d$.

One may approach this problem from the Bayesian point of view: the counts in simulation are distributed with a Poisson probability around some unknown value of the true rate $\mu$, so we convolve that probability (treating it as the likelihood) with the probability to observe the $d$ counts in data, and with some prior (taken as $\mu^z$ in the following expression):
\[
-\ln\left(\int_0^\infty {(n_d\mu)^de^{-n_d\mu}\over d!}\cdot {(n_s\mu)^se^{-n_s\mu}\over s!}\mu^zd\mu\right)=
\]
\[
-\ln{(s+d+z)!\over s! \cdot d!}-d\ln{n_d\over n_s+n_d}-s\ln{n_s\over n_s+n_d}+(z+1)\ln(n_s+n_d).
\]
For our example of section \ref{comparison} we chose $z=-1$, corresponding to a non-normalizable prior of $\mu^{-1}$.

In the next section we introduce a possible new treatment that considers statistical uncertainties in both data and simulation, and appears to perform better than the alternatives listed above, as demonstrated later in section \ref{comparison}.

\section{Likelihood description with statistical uncertainties only}
\label{llh1}
Consider a repeatable experiment that is performed $n_d$ times to collect a total of $d$ counts with a per-event expectation of $\mu_d$ (we call a single instance of this experiment an ``event''). We predict the result of the experiment with the simulation, which collects $s$ counts in $n_s$ simulated events and a per-event expectation of $\mu_s$.

Given that the total count in the combined set of simulation and data is $s+d$, the conditional probability distribution function of observing $s$ simulation and $d$ data counts is
\[P(\mu_s, \mu_d; s, d | s+d) = {(s+d)!\over s! \cdot d!} \cdot \left({n_s\mu_s\over s+d}\right)^s \cdot \left({n_d\mu_d\over s+d}\right)^d.\]
An obvious constraint that is implied here is $n_s\mu_s+n_d\mu_d=s+d$, which can be derived from the normalization requirement
\[\sum_{s,d}P(\mu_s, \mu_d; s, d | s+d)= \left({n_s\mu_s\over s+d} + {n_d\mu_d\over s+d}\right)^{s+d}=1.\]

If the data and simulation are completely unrelated the best possible estimates of $\mu_s$ and $\mu_d$ are determined by maximization of the probability function given above with the constraint $n_s\mu_s+n_d\mu_d=s+d$, which results in the estimates
\[\mu_s={s\over n_s}, \quad \mu_d={d\over n_d}.\]

Now, the alternative hypothesis that we could try to verify is that data and simulation counts are described by the same process, i.e., come with the same mean $\mu=\mu_s=\mu_d$. This identity together with the constraint $n_s\mu_s+n_d\mu_d=s+d$ uniquely determines the values of
\[\mu=\mu_s=\mu_d={s+d\over n_s+n_d}.\]

We can now compare the probabilities of the two of the above hypotheses by forming a likelihood ratio
\[
{P({\rm same\ process})\over P({\rm independent\ processes})}=\left({n_s\over n_s+n_d} / {s\over s+d}\right)^s \cdot \left({n_d\over n_s+n_d} / {d\over s+d}\right)^d=\left(\mu\over s/n_s\right)^s\cdot\left(\mu\over d/n_d\right)^d.
\]
This is the expression that we propose for comparison of different simulation sets with data. The denominator effectively factors out the dependence of the probability $P$ on the number of ``states'' (i.e., possible values of $s$) around the most likely value of $s$. (This dependence comes in because of the normalization condition: the sum of $P$ over all states should be 1, which means that the value of $P$ for the most likely values of $s$ is lower for larger $s+d$.) The expression above can also be derived starting with the Poisson probability
\[
P(\mu_s, \mu_d; s, d) = {(n_s\mu_s)^s e^{-n_s\mu_s}\over s!} \cdot {(n_d\mu_d)^d e^{-n_d\mu_d}\over d!}.
\]

\section{Generalization to many bins}
\label{llh}
If there are several bins $\{i\}$ in which simulation and data counts are compared, the conditional probability can be written as
\[P(\{\mu_s^i\}, \{\mu_d^i\}; \{s_i\}, \{d_i\} | S+D) = {(S+D)!\over \prod_i s_i! \cdot \prod_i d_i!} \cdot \prod_i \left({n_s\mu_s^i\over S+D}\right)^{s_i} \cdot \prod_i \left({n_d\mu_d^i\over S+D}\right)^{d_i}.\]
Here we use notations $S=\sum_i s_i$, $D=\sum_i d_i$. The probability sum of 1 requires $\sum_i (n_s\mu_s^i+n_d\mu_d^i)=S+D$. Taking the negative logarithm, losing constant terms, and introducing a Lagrange multiplier term for this constraint (with a new unknown $\zeta$), this becomes:
\[F= -\sum_i s_i\ln(\mu_s^in_s)-\sum_i d_i\ln(\mu_d^in_d)+\zeta\cdot(\sum_i n_s\mu_s^i+\sum_i n_d\mu_d^i-S-D).\]

If data and simulation are independent, this expression is minimized for each $\mu_s^i$, $\mu_d^i$ independently:
\[
{\partial F\over\partial \mu_s^i}=-{s_i\over\mu_s^i}+\zeta\cdot n_s=0, \quad {\partial F\over\partial \mu_d^i}=-{d_i\over\mu_d^i}+\zeta\cdot n_d=0  \quad \Rightarrow \quad \mu_s^i={s_i\over \zeta \cdot n_s}, \quad \mu_d^i={d_i\over \zeta\cdot n_d}.
\]
Plugging this back into the constraint equation (which we also get back by setting $\partial F/\partial\zeta=0$), we get
\[
\sum_i n_s {s_i\over \zeta \cdot n_s} +n_d {d_i\over \zeta\cdot n_d}=\sum_i {n_s + n_d \over\zeta}=S+D \quad \Rightarrow \quad \zeta=1.
\]

If data and simulation come from the same distribution we require $\mu^i=\mu_s^i=\mu_d^i$, and minimize against $\mu^i$:
\[
{\partial F\over\partial \mu^i}=-{s_i+d_i\over\mu^i}+\zeta\cdot (n_s+n_d)=0 \quad \Rightarrow \quad \mu^i={s_i+d_i\over \zeta\cdot (n_s+n_d)}.
\]
Once again, plugging this back into the constraint relation we get
\[
\sum_i (n_s+n_d) {s_i+d_i\over \zeta \cdot (n_s+n_d)} =\sum_i {n_s + n_d \over\zeta}=S+D \quad \Rightarrow \quad \zeta=1.
\]

Thus, the expressions derived in the previous section for 1-bin situation are valid per-bin when there are more than one bin, and we get back the likelihood ratio formula
\[
{P({\rm same\ process})\over P({\rm independent\ processes})}=\prod_i\left(\mu^i\over s_i/n_s\right)^{s_i}\cdot\prod_i\left(\mu^i\over d_i/n_d\right)^{d_i}, \quad {\rm with} \quad \mu^i={s_i+d_i\over n_s+n_d}.
\]
We compare the performance of reconstruction using this formula with the other approaches listed in the introduction in section \ref{comparison}.

\section{Likelihood description: adding model errors}
\label{sllh}

The error in describing data with simulation (i.e., describing $\mu_d$ with $\mu_s$) is often non-zero. In such a case one may quantify the amount of disagreement between data and simulation with a $\chi^2$:
\[\chi^2={(\ln\mu_d - \ln\mu_s)^2 \over \sigma^2 }.\]
Instead of setting $\mu_s=\mu_d$ as in the previous sections we assume that a difference between $\mu_s$ and $\mu_d$ can exist due to this systematic error and is modeled with a likelihood penalty term
\[\exp{\ln^2(\mu_d/\mu_s) \over -2 \sigma^2}.\]

The likelihood ratio is therefore determined as
\[
{P({\rm same\ process})\over P({\rm independent\ processes})}=\left(\mu_s\over s/n_s\right)^s\cdot\left(\mu_d\over d/n_d\right)^d \cdot \exp{\ln^2(\mu_d/\mu_s) \over -2 \sigma^2},
\]
where the $\mu_s$ and $\mu_d$ are determined by maximizing
\[P(\mu_s, \mu_d; s, d | s+d) = {(s+d)!\over s! \cdot d!} \cdot \left({n_s\mu_s\over s+d}\right)^s \cdot \left({n_d\mu_d\over s+d}\right)^d \cdot \exp{\ln^2(\mu_d/\mu_s) \over -2 \sigma^2}\]
with the constraint $n_s\mu_s+n_d\mu_d=s+d$. Taking the negative logarithm, losing constant terms, and introducing a Lagrange multiplier term for this constraint (with a new unknown $\zeta$), this becomes:
\[-s\ln(\mu_sn_s)-d\ln(\mu_dn_d) + {1 \over 2\sigma^2} \ln^2 {\mu_d \over \mu_s}+\zeta\cdot(n_s\mu_s+n_d\mu_d-s-d)\equiv F.\]

The function $F(\mu_s, \mu_d)$ can be easily minimized against $\mu_s$ and $\mu_d$, yielding estimates of these quantities. To demonstrate this, first the derivatives of $F$ are calculated and set to 0:
\[\mu_s {\partial F \over \partial \mu_s} = \zeta\mu_sn_s - s - {1 \over \sigma^2} \ln{\mu_d \over \mu_s} = 0,\]
\[\mu_d {\partial F \over \partial \mu_d} = \zeta\mu_dn_d - d + {1 \over \sigma^2} \ln{\mu_d \over \mu_s} = 0.\]
The sum of these, $\zeta\cdot(\mu_sn_s + \mu_dn_d) = s+d$, results in the value for $\zeta=1$. The derivative of $F$ with respect to $\zeta$ gives back the constraint $n_s\mu_s+n_d\mu_d=s+d$, which yields an expression of $\mu_d$ as a function of $\mu_s$. Plugging it into the first of the above two equations one gets
\[f=\mu_s {\partial F \over \partial \mu_s} (\mu_s, \mu_d(\mu_s)) = \mu_sn_s - s - {1 \over \sigma^2} \ln{\mu_d(\mu_s) \over \mu_s} = 0.\]

This equation can be solved with a few iterations of the Newton's root finding method starting with a solution to
\[\mu_s=\mu_d(\mu_s)\mbox{:} \quad \quad \mu_s = \mu_d = {s+d \over n_s+n_d}.\]
At each iteration the value of $\mu_s$ is adjusted by $-f/f^\prime$, where the derivative is evaluated as
\[f^\prime=n_s\left( 1+{1 \over \sigma^2} ({1 \over \mu_sn_s} + {1 \over \mu_dn_d})\right).\]

Once the likelihood function is solved for the best values of $\mu_s$ and $\mu_d$, these can be plugged into the likelihood ratio given above. One can now write the likelihood ratio as a sum over all bins:
\[-\ln{\cal L}=\sum_i{\left[ s_i\ln{s_i/n_s\over\mu_s^i}+d_i\ln{d_i/n_d\over\mu_d^i}+{1\over 2\sigma^2}\ln^2{\mu_d^i\over\mu_s} \right] }.\]
This is an improved expression compared to the one used in \cite{ice}, and has been applied in an updated analysis of \cite{lea}. The probability $P({\rm same\ process})$ can also be thought of as a convolution of the binomial probability part of the expression with the penalty term. Solving the convolution integral approximately with the Laplace's method results (up to a constant term) in an expression for $P({\rm same\ process})$ given above.

\section{Likelihood description of data with weighted simulation}
\label{wllh}
One can apply the method for calculating the likelihood ratio of the previous section to a situation that is common when the number of data counts $d_k$ in bin $k$ measured during time $t_d$ is fitted with a number of simulation counts $s_{ki}$, each representing a possibly different (for weighted simulation) time $t_{ki}$ (usually related to the event weight $w_{ki}$ as $w_{ki}\cdot t_{ki}=t_d$). Although we can assume that all $s_{ki}=1$ without the loss of generality, we continue with the notation $s_{ki}$. The combined number of events in data and simulation is then $S+D$, where $S=\sum_k s_k$, $s_k=\sum_i s_{ki}$, $D=\sum_k d_k$. The expression for the conditional probability is now
\[P(\{\mu_s^{ki}\}, \{\mu_d^k\}; \{s_{ki}\}, \{d_k\} | S+D) = {(S+D)!\over \prod_{ki} s_{ki}! \cdot \prod_k d_k!} \cdot \prod_{ki} \left({t_{ki}\mu_s^{ki}\over S+D}\right)^{s_{ki}} \cdot \prod_k\left({t_d\mu_d^k\over S+D}\right)^{d_k}.\]

The probability sum of 1 necessitates the constraint
\[\sum_{ki} t_{ki}\mu_s^{ki}+\sum_k t_d\mu_d^k=S+D.\]
Taking the negative logarithm of $P$, losing constant terms, and introducing a Lagrange multiplier term for this constraint (with a new unknown $\zeta$), we get:
\[
F=-\sum_{ki}s_{ki}\ln(t_{ki}\mu_s^{ki})-\sum_kd_k\ln(t_d\mu_d^k)+\zeta\cdot\left( \sum_{ki} t_{ki}\mu_s^{ki}+\sum_k t_d\mu_d^k-S-D \right)
\]

If data and simulation are independent, this expression can be minimized for each $\mu_s^{ki}$, $\mu_d^k$ independently:
\[
{\partial F\over \partial\mu_s^{ki}}=-{s_{ki}\over\mu_s^{ki}}+\zeta\cdot t_{ki}=0, \quad {\partial F\over \partial\mu_d^k}=-{d_k\over \mu_d^k}+\zeta\cdot t_d=0 \quad \Rightarrow \quad \mu_s^{ki}={s_{ki}\over\zeta\cdot t_{ki}}, \quad \mu_d^k={d_k\over\zeta\cdot t_d}.
\]
Plugging this back into the constraint equation (which we also get back by setting $\partial F/\partial\zeta=0$), we get
\[
\sum_{ki}t_{ki}{s_{ki}\over\zeta\cdot t_{ki}}+\sum_k t_d{d_k\over\zeta\cdot t_d}=\sum_k{s_k+d_k\over\zeta}=S+D \quad \Rightarrow \quad \zeta=1.
\]

If data and simulation come from the same distribution we require $\mu_d^k=\sum_i \mu_s^{ki}$ for each $k$. These conditions can be introduced into the above expression for $F$ as additional terms (with new unknowns $\xi_k$):
\[
F=-\sum_{ki}s_{ki}\ln(t_{ki}\mu_s^{ki})-\sum_kd_k\ln(t_d\mu_d^k)+\zeta\cdot\left( \sum_{ki} t_{ki}\mu_s^{ki}+\sum_k t_d\mu_d^k-S-D \right)+\sum_k\xi_k\cdot\left(\sum_i \mu_s^{ki}-\mu_d^k\right).
\]
Derivatives with respect to $\zeta$ and $\xi_k$ give back the constraint equations. The other derivatives are:
\[
{\partial F\over\partial\mu_s^{ki}}=-{s_{ki}\over\mu_s^{ki}}+\zeta\cdot t_{ki}+\xi_k=0, \quad {\partial F\over\partial\mu_d^k}=-{d_k\over\mu_d^k}+\zeta\cdot t_d-\xi_k=0.
\]
Multiplying the first equation by $\mu_s^{ki}$, the second by $\mu_d^k$, and summing them together, we get
\[
0=-\sum_{ki} s_{ki}-\sum_k d_k+\zeta\cdot\left(\sum_{ki}t_{ki}\mu_s^{ki}+\sum_kt_d\mu_d^k\right)+\sum_k\xi_k\cdot\left(\sum_i\mu_s^{ki}-\mu_d^k\right)=-S-D+\zeta\cdot(S+D)+\sum_k\xi_k\cdot 0.
\]
Therefore $\zeta=1$. To find $\xi$ we substitute the expressions for $\mu_s^{ki}$ and $\mu_d^k$ into constraints for $\xi_k$:
\[
\mu_s^{ki}={s_{ki}\over t_{ki}+\xi_k}, \quad \mu_d^k={d_k\over t_d-\xi_k} \quad \Rightarrow \quad \sum_i {s_{ki}\over t_{ki}+\xi_k} = {d_k\over t_d-\xi_k}.
\]

Therefore, the likelihood ratio is
\[
{P({\rm same\ process})\over P({\rm independent\ processes})}=\prod_{ki}\left(t_{ki}\over t_{ki}+\xi_k\right)^{s_{ki}}\cdot \prod_k\left(t_d\over t_d-\xi_k\right)^{d_k}.
\]

The equation for $\xi_k$ is similar to equation 15 of \cite{barlow}. As suggested there, we solve them for each $k$ by Newton's method starting with $\xi_k=0$, ensuring that $-\min_{\{s_{ki}>0\}}(t_{ki})<\xi_k\leq t_d$:
\[
f_k=1/\left[\sum_i {s_{ki}\over t_{ki}+\xi_k}\right]-{t_d-\xi_k\over d_k}, \quad {df_k\over d\xi_k}=\sum_i {s_{ki}\over (t_{ki}+\xi_k)^2}/\left[\sum_i {s_{ki}\over t_{ki}+\xi_k}\right]^2+{1\over d_k} \quad \Rightarrow \quad \xi_k({\rm next})=\xi_k-{f_k\over df_k/d\xi_k}.
\]
The particular form of function $f_k$ above (inverted compared to the original equation for $\xi_k$) was chosen to linearize the problem in simple cases (e.g., all simulated events having the same weight). After the first iteration (starting with $\xi_k=0$) we get
\[
\xi_k\approx{t_d/d_k-1/\sum_is_{ki}/t_{ki}\over 1/d_k+\sum_is_{ki}/t_{ki}^2/[\sum_is_{ki}/t_{ki}]^2}=t_d\cdot{1-d_k/m_k\over 1+d_k\cdot\varepsilon_k^2/m_k^2},
\]
\[{\rm with} \quad m_k=\sum_i s_{ki}w_{ki}, \quad \varepsilon_k^2=\sum_i s_{ki}w_{ki}^2, \quad w_{ik}={t_d\over t_{ik}}.
\]
Here $m_k$ is the total simulation in bin $k$ evaluated for time $t_d$ (i.e., is a prediction of data in bin $k$); $\varepsilon_k$ is a statistical uncertainty of the value $m_k$. If $m_k$ and $d_k$ are not too far from each other, and the statistical uncertainty of the simulation is much smaller than that of data, i.e., $\varepsilon_k\ll\sqrt{d_k}$, then $\xi_k\approx t_d\cdot(1-d_k/m_k)$. This holds in the limit of infinite simulation statistics, when all $t_{ki}\rightarrow\infty$, and thus, $w_{ki}\rightarrow 0$. Continuing with this approximation,
\[
{P({\rm same\ process})\over P({\rm independent\ processes})}=\prod_{ki}\left(1\over 1+w_{ki}\cdot(1-d_k/m_k)\right)^{s_{ki}}\cdot \prod_k\left(1\over 1-(1-d_k/m_k)\right)^{d_k}\approx
\]
\[
\exp\left[-\sum_{ki}s_{ki}w_{ki}\cdot(1-d_k/m_k)\right]\cdot \prod_k\left(m_k\over d_k\right)^{d_k}=\prod_k \left[\exp(d_k-m_k)\cdot\left(m_k\over d_k\right)^{d_k}\right]
\]

Up to a constant term this is a product of the usual expressions for the Poisson likelihood,
\[
p(m; d)={m^d\over d!}\cdot\exp(-m).
\]
However, the expression above is also precisely identical to the product of ratios $p(m; d)/p(d; d)$, where $m=d$ in the denominator is the value at which the Poisson likelihood $p(m; d)$ achieves its maximum.

We would like to emphasize that using the usual Poisson expression for the likelihood fit of data to simulation is, strictly speaking, only correct in the limit of counts of the simulation being infinite in all of the bins used in the likelihood fit, and all of the simulation event weights being infinitesimally small. An uneven distribution of simulation counts in bins (e.g., energy bins) may create a bias in the resulting fit (i.e., bias in energy). In this section we advocate using the exact expression for the likelihood ratio given above. We are aware of several recent works (e.g., \cite{anne}) trying the new expression and finding that it works better than the usual Poisson likelihood.

\section{Likelihood description with weighted simulation and model errors}
\label{swllh}
Instead of assuming the condition of $\sum_i \mu_s^{ki}=\mu_d^k$ used in section \ref{wllh} we add a penalty term similar to section \ref{sllh}:
\[
\prod_k\exp\left(-{1\over 2\sigma^2}\ln^2{\sum_i \mu_s^{ki}\over\mu_d^k}\right).
\]

To simplify the following expressions we introduce notations $\mu_s^k=\sum_i \mu_s^{ki}$ and $\xi_k=\ln(\mu_s^k/\mu_d^k)$. So, if the data and simulation come from the similar distributions, discrepancies being described by the above penalty term, the expression for F is:
\[
F=-\sum_{ki}s_{ki}\ln(t_{ki}\mu_s^{ki})-\sum_kd_k\ln(t_d\mu_d^k)+\zeta\cdot\left( \sum_{ki} t_{ki}\mu_s^{ki}+\sum_k t_d\mu_d^k-S-D \right)+{1\over 2\sigma^2}\sum_k\xi_k^2.
\]
The derivatives with respect to the unknown $\mu_s^{ki}$, $\mu_d^k$ are:
\[
{\partial F\over\partial\mu_s^{ki}}=-{s_{ki}\over\mu_s^{ki}}+\zeta\cdot t_{ki}+{\xi_k\over\sigma^2\mu_s^k}=0, \quad {\partial F\over\partial\mu_d^k}=-{d_k\over\mu_d^k}+\zeta\cdot t_d-{\xi_k\over\sigma^2\mu_d^k}=0.
\]
As in section \ref{wllh} we prove that $\zeta=1$ by multiplying these equations by $\mu_s^{ki}$ and $\mu_d^k$ and summing them together. We can rewrite the above equations to get expressions for $\mu_s^{ki}$ and $\mu_d^k$:
\[
\mu_s^{ki}={s_{ki}\over t_{ki}+{\xi_k\over\sigma^2\mu_s^k}}, \quad \mu_d^k={1\over t_d}\left(d_k+{\xi_k\over\sigma^2}\right).
\]
If we find $\xi_k$ we could determine all of $\mu_s^{ki}$ and $\mu_d^k$: first, $\mu_d^k$ is given by the second of the above equations, then we use $\mu_s^k=\xi_k\mu_d^k$ and determine $\mu_s^{ki}$ from the first of the above equations. We can now write an equation for $\xi_k$:
\[
\mu_s^k=\sum_i \mu_s^{ki}=\sum_i{s_{ki}\over t_{ki}+{\xi_k\over\sigma^2\mu_s^k}} \quad \Rightarrow \quad \sum_i{s_{ki}\over \mu_s^k t_{ki}+{\xi_k/\sigma^2}}=1, \quad \mbox{where} \quad \mu_s^k={e^{\xi_k}\over t_d}\left(d_k+{\xi_k\over\sigma^2}\right).
\]
Similar to section \ref{wllh}, this equation can be solved with Newton's method starting with $\xi_k=1$. We can now write the likelihood ratio as
\[
{P({\rm similar\ process})\over P({\rm independent\ processes})}=\prod_{ki}\left({t_{ki}\mu_s^{ki}\over s_{ki}}\right)^{s_{ki}}\cdot \prod_k\left({t_d\mu_d^k\over d_k}\right)^{d_k}\cdot\prod_k\exp\left(-{1\over 2\sigma^2}\xi_k^2\right).
\]
To find optimal parameters of the simulation one typically maximizes this by minimizing the minus log likelihood:
\[
-\ln{\cal L}=\sum_{ki}s_{ki}\ln\left({s_{ki}\over t_{ki}\mu_s^{ki}}\right) + \sum_kd_k\ln\left({d_k\over t_d\mu_d^k}\right) +{1\over 2\sigma^2} \sum_k\xi_k^2.
\]
We note that the derivatives of this expression with respect to parameters $t_{ki}$ are equal to the partial derivatives of this expression with respect to parameters $t_{ki}$, since the rest has terms proportional to the derivatives with respect to $\mu_s^{ki}$ and $\mu_d^k$, which were set to 0 by the above effort. This could be useful when searching for the optimal values of $t_{ki}$ themselves.

\section{Adding noise}
The likelihood expressions of the previous two sections can only be computed when all bins that have data counts contain at least one or more simulated counts. If there are bins with data but no simulation there must have been some effect that was not simulated, or the amount of simulation was insufficient. In such cases the likelihood calculation is impossible unless one excludes these bins from the consideration. For a meaningful comparison of likelihood ratio values computed for different simulation sets all bins that have no simulated counts in at least one of the simulated sets must be excluded from all likelihood evaluations.

\begin{figure}
\begin{center}
	\includegraphics[width=0.5\linewidth]{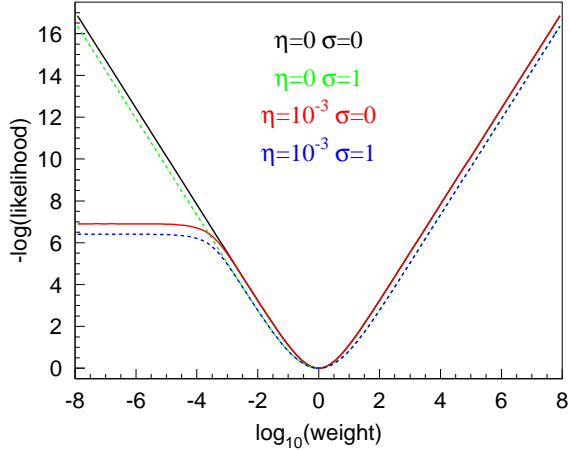}
	\caption{Dependence of the one-term likelihood on the simulation weight. The data and simulation are both set to 1 count, so the minimum is at $w=1$, as expected. Without noise the likelihood expressions of both sections \ref{wllh} ($\sigma=0$) and \ref{swllh} ($\sigma=1$) diverge at 0 (i.e., at large negative values of $\log_{10}w$). Adding noise (with rate $\eta=10^{-3}$) results in a finite value at 0 for both likelihood expressions. Also note that the curves with $\sigma=1$ have wider minima, as expected.}
	\label{ns}
\end{center}
\end{figure}

It is sometimes desired to not simulate a simple effect of a known rate $\eta_k$ (i.e., noise) but rather directly introduce $\eta_k$ into the likelihood calculation. This would coincidentally solve the problem with bins that have no simulated counts described in the preceding paragraph. The constraint for rates of section \ref{wllh} is modified as $\sum_i \mu_s^{ki}+\eta_k=\mu_d^k$, and the penalty term of section \ref{swllh} is modified via a new expression for $\xi_k$: $\xi_k=\ln[(\mu_s^k+\eta_k)/\mu_d^k]$. When the value $\eta_k$ of noise rate is known much better than the $\sigma$ or the penalty term assumes, the new expression for $\xi_k$ may be written as $\xi_k=\ln[\mu_s^k/(\mu_d^k-\eta_k)]$ instead. Although the expression for $\zeta$ becomes more complicated, it can be still approximated as 1 in the case of many bins with a comparably large total number of counts $S+D$. Figure \ref{ns} compares the behavior of the likelihood functions (with only one simulation and one data counts) of sections \ref{wllh} and \ref{swllh} with and without noise.

\section{Binning}
\label{binning}

If the data is binned in equal-length consecutive bins one may occasionally benefit from combining the counts in contiguous subsets of bins into larger bins of length L (L is the number of initial-size small bins in the larger bin). The benefit of combining the counts of several bins must be weighted against the possibility of measuring the change in rate between the bins. If the counts are comparable to each other or are very small in the two adjacent bins (i.e., are within the statistical uncertainties), then the rate change across the these bins is consistent with zero, and thus the bins should be combined. This is the principal idea behind the ``Bayesian blocks'' algorithm of \cite{scargle}. Below we summarize the algorithm, as it is used in section \ref{comparison}. Note how the probability ratio construction used by this algorithm is quite similar to the probability ratio used in this paper.

The multinomial probability for a particular distribution of counts in bins under consideration is
\[P(\{\mu_i\}; \{d_i\} | \sum_id_i=D) = {D!\over\prod_i d_i!} \cdot \mu_i^{d_i}.\]

The first hypothesis is that the counts in these bin are independent from each other. It follows that $\mu_i=d_i$ for all $i$. The second hypothesis is that the counts in all of the bins under consideration are governed by the process with the same rate $\mu=\mu_i$ for all $i$. If follows that $\mu=D/L$, where L is the number of bins.

To determine whether the bins should be combined or be considered separately we form the likelihood ratio:
\[
{P({\rm one\ bin})\over P({\rm many\ bins})}={(D/L)^D\over \prod_i d_i^{d_i} }, \quad {\rm where} \quad D=\sum_i d_i.
\]

Now we note that this ratio never exceeds 1, as the value of the likelihood maximized with a constraint $\mu=\mu_i$ is necessarily smaller or equal to than the value of the likelihood maximized without any restrictions.

Note that for any permutation of bins $i$ the ratio above does not change. However, with the assumption of the same rate across the bins all ($L!$) of the permutations are equivalent to merely being different statistical realizations of the same experiment, all leading to the same term $\prod_i d_i^{d_i}$ in the denominator. On the other hand there is only one possible realization of the data arrangement that leads to the term $(D/L)^D$ in the nominator.

We will now make an intuitive conclusion that it is fair to decide whether to combine the bins $i$ into a single large bin with length $L$ based on a comparison of the ratio above with $1/L!$. We can generalize this rule for comparing various arrangements of bins $k$ of different sizes $L_k$: we choose the arrangement that leads to the largest sum (taking the log of the probability ratio)
\[
\sum_k D_k\cdot \ln\left(D_k\over L_k\right) + \ln(L_k!).
\]
This is different from the consideration given in \cite{scargle} in the expression for the last term, which in \cite{scargle} is given by $\approx -\ln(8)$. This reference also contains a description of the algorithm for the search of the best bin arrangement, which is easy to implement and is very efficient.

To get a little more insight into the expression given above we write it for a number $L$ of bins $i$ each with approximately the same counts $d_i=d+\delta_i$, $D=\sum_i d_i = d\cdot L$, $\delta_i/d\ll 1$:
\[
\sum_i d_i\cdot \ln d_i + \ln(1!)=\sum_i(d+\delta_i)\ln(d+\delta_i) \approx \sum_i \left(d\ln d +{\delta_i^2\over 2d}\right)=D\ln\left({D\over L}\right) +{1\over 2}\sum_i{\delta_i^2\over d}.
\]

We compare this with the expression for a single large bin of length $L$:
\[
D\ln\left({D\over L}\right)+\ln(L!).
\]

So, the decision to combine the bins is based on whether
\[
{1\over 2}\sum_i{\delta_i^2\over d}<\ln(L!)\approx L\cdot\ln\left({L\over e}\right)\sim L, \quad {\rm or} \quad \sum_i{\delta_i^2\over d}\lesssim L,
\]
i.e. when the $\chi^2$ (i.e., goodness-of-fit) of description of data with the mean $d$ in the bins is on the order of or smaller than the number of bins (i.e., degrees of freedom of the $\chi^2$).

\section{Comparison with other methods}
\label{comparison}
In this section we will compare the performance of parameter reconstruction using the likelihood ratio expression introduced in section \ref{llh} and three approaches enumerated in the introduction: Poisson, $\chi^2$, and Bayesian. We also reconstruct using the ``true'' Poisson likelihood by calculating the expectation in each bin directly from the probability distribution function (PDF). We take the distribution of counts following the following PDF: $\exp(-x/\mu)/\mu$ with $\mu_0=5$. The sets are drawn with the mean total number of counts per set of 10. This ``experiment'' is performed $n_d=1$, 10, 100, or 1000 times, and each simulation is performed $n_s=1$, 10, 100, 1000 times.

\begin{figure}[h!]
\begin{center}
	\includegraphics[width=0.45\linewidth]{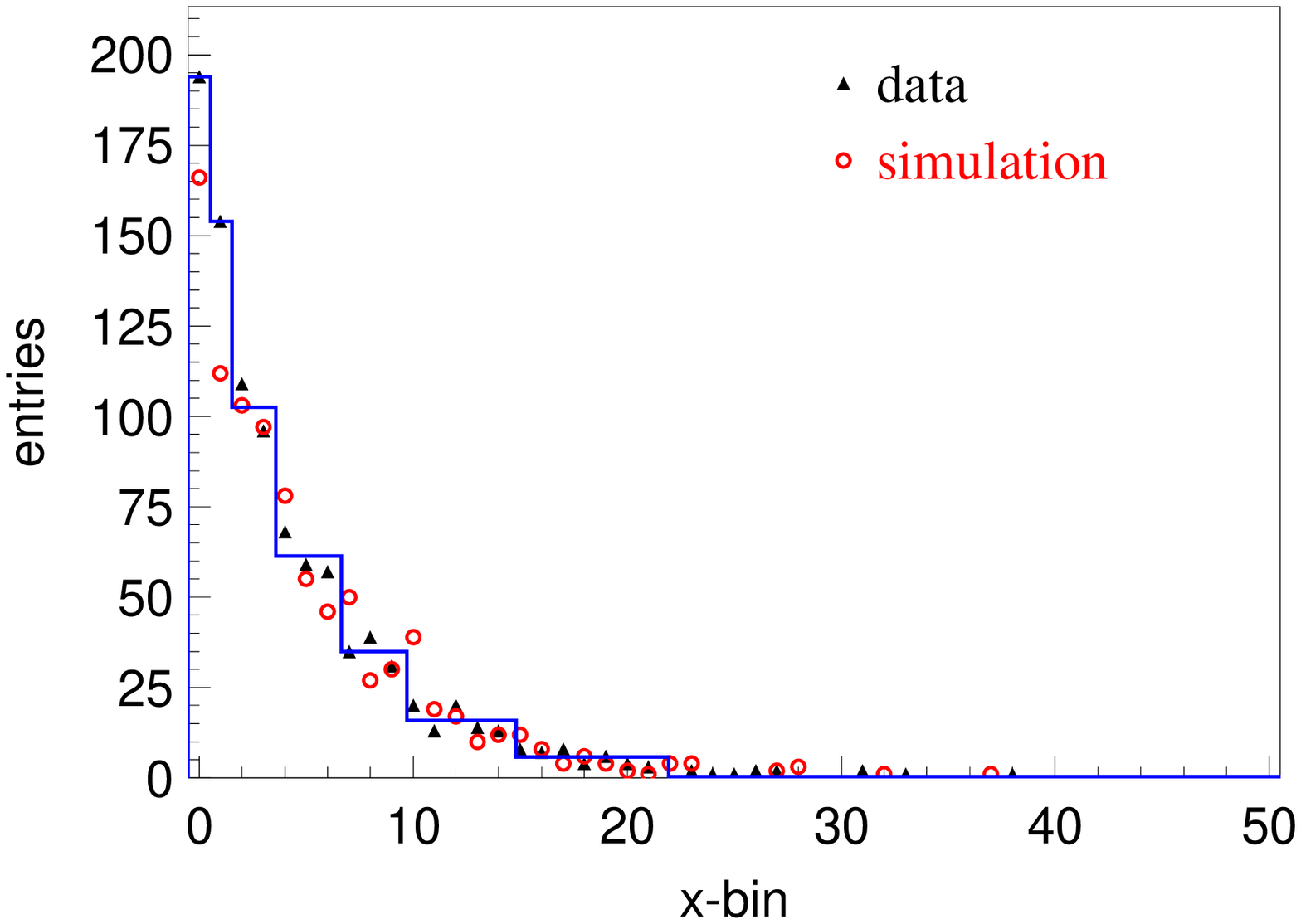}
	\includegraphics[width=0.45\linewidth]{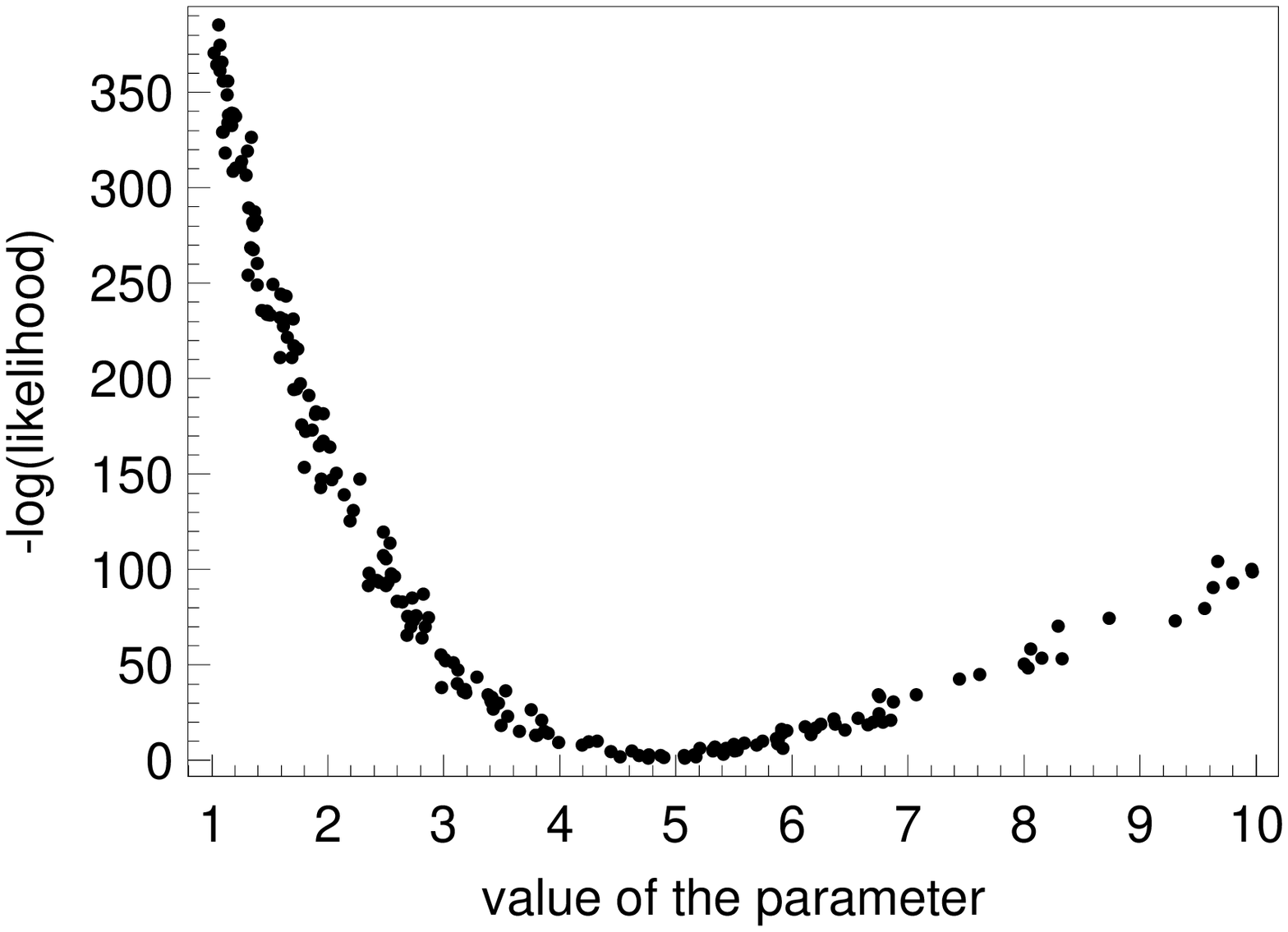}
	\caption{Left: data and simulation for $n_d=100$ and $n_s=100$. Also shown in blue solid line is the binning optimized with the method described in section \ref{binning}. Right: likelihood ratio values for random simulations in the vicinity of the true value of $\mu_0=5$.}
	\label{exa}
\end{center}
\end{figure}

The data and simulation are optionally rebinned according to the method described in section \ref{binning}. We investigated the binning strategy optimized for combined counts of data and simulation, which often gave the best result, and compared it to the binning optimized for counts of just data, which is often almost as good but more practical (as it is only done once per data sample). The reconstruction of the value of parameter $\mu$ is performed by simulating at points randomly and log-uniformly sampled in the range $\mu_0/5$ to $2\mu_0$ and comparing resulting likelihood ratio values. A typical result of such sampling procedure is shown in Figure \ref{exa} along with a comparison between a representative data sample and a simulation sample at the best fit point.

\begin{figure}[h!]
\begin{center}
	\includegraphics[width=\linewidth]{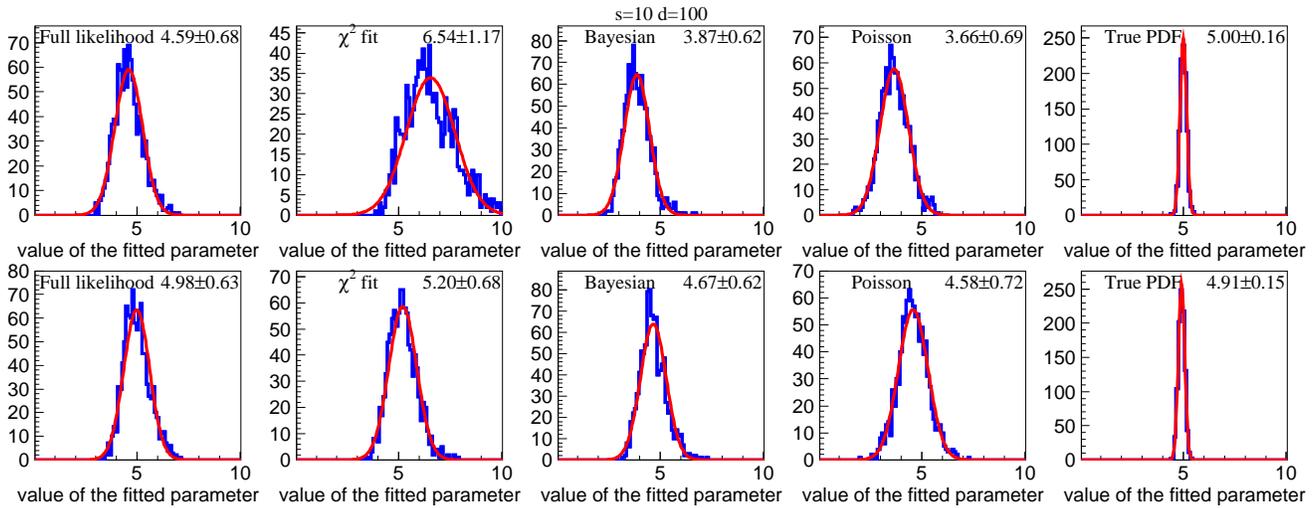}
	\caption{Distributions of the reconstructed value of $\mu$ for 1000 random drawings of data with $n_d=100$. For each data set, 200 simulation sets with $\mu$ sampled log-uniformly in the interval $(\mu_0/5,\ 2\mu_0)$ are drawn with $n_s=10$ and the one with the best likelihood value is chosen to represent the result of the fit. The top row uses the bins of size 1, the bottom row uses the bins with size optimized according to the method of section \ref{binning}. The Poisson ``true PDF'' likelihood is always the best, as expected, and does not depend on the binning method. In all other cases the result is improved with the optimized binning. The ``full likelihood'' corresponds to the likelihood description of this paper and gives the best result in this example.}
	\label{all}
\end{center}
\end{figure}

We compare our findings in Table \ref{res}. A representative set of distributions of the reconstructed values of $\mu$ is shown in Figure \ref{all} for $s=10$, and $d=100$. The new expression gives the best result (closest to $\mu_0=5$ with smallest spread) more often than the other methods, except, of course, the method using the Poisson likelihood with the mean estimated from the analytical expression for the PDF. We also comment that the optimized binning improves the result in most cases.

\begin{table}[h!]
\begin{center}
\begin{tabular}{|cc|ccccc|}
\hline
\multicolumn{2}{|c}{} & \multicolumn{5}{|c|}{default binning} \\
$n_s$ & $n_d$ & Full & $\chi^2$ & Bayesian & Poisson & True PDF \\
1 & 1 & { 3.79 $\pm$ 1.97} & { 3.82 $\pm$ 1.98} & {\bf 5.71 $\pm$ 2.80} & { 3.06 $\pm$ 2.30} & { 5.02 $\pm$ 1.71} \\
1 & 10 & { 3.53 $\pm$ 1.52} & {\bf 5.15 $\pm$ 1.80} & {\it 2.82 $\pm$ 1.98} & {\it 2.68 $\pm$ 1.88} & { 5.00 $\pm$ 0.51} \\
1 & 100 & {\it 3.46 $\pm$ 1.45} & {\bf 5.66 $\pm$ 1.88} & {\it 2.35 $\pm$ 1.44} & {\it 2.50 $\pm$ 1.64} & { 5.00 $\pm$ 0.16} \\
1 & 1000 & {\it 3.39 $\pm$ 1.43} & {\bf 5.61 $\pm$ 1.90} & {\it 2.31 $\pm$ 1.33} & {\it 2.47 $\pm$ 1.57} & { 5.00 $\pm$ 0.07} \\
10 & 1 & {\bf 4.48 $\pm$ 1.80} & {\it 2.06 $\pm$ 1.37} & { 6.16 $\pm$ 2.97} & {\it 1.48 $\pm$ 0.59} & { 5.09 $\pm$ 1.72} \\
10 & 10 & { 4.62 $\pm$ 0.94} & {\bf 4.63 $\pm$ 0.96} & { 3.99 $\pm$ 1.26} & {\it 2.40 $\pm$ 0.88} & { 5.00 $\pm$ 0.50} \\
10 & 100 & {\bf 4.59 $\pm$ 0.67} & {\it 6.54 $\pm$ 1.17} & {\it 3.87 $\pm$ 0.62} & {\it 3.66 $\pm$ 0.69} & { 5.00 $\pm$ 0.16} \\
10 & 1000 & {\bf 4.64 $\pm$ 0.65} & {\it 8.61 $\pm$ 0.98} & {\it 3.85 $\pm$ 0.57} & {\it 3.96 $\pm$ 0.65} & { 5.00 $\pm$ 0.06} \\
100 & 1 & {\bf 4.78 $\pm$ 1.75} & {\it 1.52 $\pm$ 1.14} & { 6.65 $\pm$ 3.11} & {\it 2.09 $\pm$ 1.21} & { 5.00 $\pm$ 1.69} \\
100 & 10 & {\bf 4.86 $\pm$ 0.65} & {\it 3.35 $\pm$ 0.51} & { 4.41 $\pm$ 0.92} & {\it 4.05 $\pm$ 0.89} & { 5.01 $\pm$ 0.52} \\
100 & 100 & { 4.94 $\pm$ 0.35} & {\bf 4.95 $\pm$ 0.34} & {\it 4.63 $\pm$ 0.34} & { 4.72 $\pm$ 0.36} & { 5.00 $\pm$ 0.16} \\
100 & 1000 & {\bf 4.93 $\pm$ 0.26} & {\it 5.39 $\pm$ 0.33} & {\it 4.71 $\pm$ 0.26} & { 4.86 $\pm$ 0.28} & { 5.00 $\pm$ 0.06} \\
1000 & 1 & {\bf 4.87 $\pm$ 1.66} & {\it 1.44 $\pm$ 1.09} & { 6.88 $\pm$ 3.21} & {\it 3.11 $\pm$ 1.77} & { 4.94 $\pm$ 1.67} \\
1000 & 10 & {\bf 4.94 $\pm$ 0.56} & {\it 2.25 $\pm$ 0.37} & { 4.51 $\pm$ 0.82} & { 4.72 $\pm$ 0.70} & { 5.01 $\pm$ 0.50} \\
1000 & 100 & {\bf 4.98 $\pm$ 0.23} & {\it 4.46 $\pm$ 0.23} & {\it 4.76 $\pm$ 0.23} & { 4.93 $\pm$ 0.24} & { 4.99 $\pm$ 0.17} \\
1000 & 1000 & {\bf 4.99 $\pm$ 0.13} & { 4.99 $\pm$ 0.13} & { 4.92 $\pm$ 0.13} & { 4.98 $\pm$ 0.13} & { 5.00 $\pm$ 0.06} \\

\hline
\multicolumn{2}{|c}{} & \multicolumn{5}{|c|}{optimized binning} \\
$n_s$ & $n_d$ & Full & $\chi^2$ & Bayesian & Poisson & True PDF \\
1 & 1 & { 3.57 $\pm$ 2.22} & { 3.57 $\pm$ 2.22} & { 3.73 $\pm$ 2.43} & {\bf 4.05 $\pm$ 2.70} & { 3.81 $\pm$ 2.44} \\
1 & 10 & {\bf 5.05 $\pm$ 1.80} & { 5.29 $\pm$ 1.72} & { 3.99 $\pm$ 1.75} & {\it 2.61 $\pm$ 1.89} & { 4.53 $\pm$ 0.49} \\
1 & 100 & {\bf 4.95 $\pm$ 1.69} & {\it 7.00 $\pm$ 1.66} & { 3.47 $\pm$ 1.65} & {\it 2.76 $\pm$ 1.81} & { 4.92 $\pm$ 0.15} \\
1 & 1000 & { 3.62 $\pm$ 1.49} & {\bf 5.86 $\pm$ 1.92} & {\it 2.30 $\pm$ 1.31} & {\it 2.51 $\pm$ 1.61} & { 4.99 $\pm$ 0.07} \\
10 & 1 & { 3.42 $\pm$ 2.13} & {\bf 3.48 $\pm$ 2.20} & { 3.46 $\pm$ 2.11} & { 3.31 $\pm$ 2.18} & { 3.71 $\pm$ 2.34} \\
10 & 10 & {\bf 4.62 $\pm$ 0.78} & { 4.62 $\pm$ 0.78} & { 4.43 $\pm$ 0.80} & {\it 3.22 $\pm$ 1.48} & { 4.53 $\pm$ 0.48} \\
10 & 100 & {\bf 4.98 $\pm$ 0.63} & { 5.20 $\pm$ 0.68} & { 4.66 $\pm$ 0.62} & { 4.58 $\pm$ 0.72} & { 4.91 $\pm$ 0.15} \\
10 & 1000 & {\bf 5.01 $\pm$ 0.71} & {\it 7.05 $\pm$ 1.14} & {\it 4.26 $\pm$ 0.62} & { 4.28 $\pm$ 0.75} & { 4.99 $\pm$ 0.06} \\
100 & 1 & { 3.48 $\pm$ 2.16} & {\bf 3.63 $\pm$ 2.35} & { 3.53 $\pm$ 2.22} & { 3.47 $\pm$ 2.17} & { 3.90 $\pm$ 2.52} \\
100 & 10 & {\bf 4.55 $\pm$ 0.56} & { 4.48 $\pm$ 0.63} & { 4.44 $\pm$ 0.57} & { 4.34 $\pm$ 0.98} & { 4.54 $\pm$ 0.50} \\
100 & 100 & {\bf 4.92 $\pm$ 0.27} & { 4.92 $\pm$ 0.27} & { 4.85 $\pm$ 0.27} & { 4.92 $\pm$ 0.26} & { 4.92 $\pm$ 0.16} \\
100 & 1000 & { 4.99 $\pm$ 0.24} & { 5.09 $\pm$ 0.25} & { 4.90 $\pm$ 0.24} & {\bf 5.00 $\pm$ 0.25} & { 4.99 $\pm$ 0.06} \\
1000 & 1 & { 3.33 $\pm$ 2.12} & {\bf 3.53 $\pm$ 2.39} & { 3.51 $\pm$ 2.32} & { 3.33 $\pm$ 2.12} & { 3.82 $\pm$ 2.46} \\
1000 & 10 & {\bf 4.52 $\pm$ 0.51} & { 4.38 $\pm$ 0.75} & {\it 4.42 $\pm$ 0.53} & { 4.49 $\pm$ 0.60} & { 4.53 $\pm$ 0.49} \\
1000 & 100 & {\bf 4.91 $\pm$ 0.20} & { 4.86 $\pm$ 0.21} & { 4.87 $\pm$ 0.19} & { 4.91 $\pm$ 0.20} & { 4.91 $\pm$ 0.16} \\
1000 & 1000 & { 4.99 $\pm$ 0.11} & { 4.99 $\pm$ 0.11} & { 4.97 $\pm$ 0.11} & {\bf 5.00 $\pm$ 0.11} & { 4.99 $\pm$ 0.06} \\

\hline
\end{tabular}
\caption{Values of $\mu$ reconstructed by various methods. Intervals are shown with mean and rms of the reconstructed values for 1000 of random data drawings. The result of the first four methods (i.e., except Poisson with true PDF shown in the last column) closest to $\mu_0=5$ is shown with bold font. For all five methods intervals that do not enclose the true value of $\mu_0=5$ (i.e., those that are significantly biased) are shown with italic font. The column designated with ``Full'' corresponds to the likelihood description of this paper, section \ref{llh}.}
\label{res}
\end{center}
\end{table}

\section{Conclusion}
In this paper we consider a problem of describing data with limited-statistics simulation sets. After a brief review of the usual methods we presented an alternative approach to the likelihood-based comparison of data with multiple simulation sets. In our tests the new approach appears to improve the results of model parameter fits, mainly reducing the bias and uncertainty of the fitted parameters. We remark that depending on the statistics of the data and available simulation, and on the nature of the experiment other approaches may occasionally produce a better result, so we encourage testing them as well, as we did in section \ref{comparison}.

The method of this paper as described in sections \ref{llh1}-\ref{sllh} was developed as part of calibration work reported in \cite{ice}, \cite{lea}. The extension to weighted simulation described in section \ref{wllh} was proposed for use in several analyses of IceCube, and was shown to yield slightly better results (e.g., in \cite{anne}: better limits, due to narrower distributions of test statistic). The method of section \ref{swllh} was recently applied to the problem of unfolding energy losses along the muon track in ice with direct re-simulation \cite{dirf}; the correct account of the model error term is important since the model error was not completely removed by our calibration efforts (see \cite{ice}, \cite{lea}).

\section*{Acknowledgment}
This work was supported by the U.S. National Science Foundation-Office of Polar Programs and the U.S. National Science Foundation-Physics Division. I would also like to thank Segev BenZvi, Markus Ahlers, Gary Hill, Chris Weaver, and Klas Hultqvist for useful discussions, and Anne Schukraft, Marius Wallraff, Sebastian Euler, Laura Gladstone, and Juan Pablo Yanez for early adoption of the method and encouragement to proceed with the publication.

\end{document}